\documentclass[11pt]{report}
\usepackage{graphicx}
\usepackage[nointegrals]{wasysym}
\usepackage{amsmath}
\usepackage{amsfonts}
\usepackage{hyperref}
\usepackage{url}
\usepackage{mathtools}
\usepackage[margin=0.9in]{geometry}
\usepackage[numbers,sort&compress]{natbib}
\usepackage{float}
\hypersetup{
	pdftitle={Report PDF Name},
	pdfauthor={Thomas Swain},
	pdfsubject={Report Subject},
	pdfkeywords={keyword1, keyword2},
	bookmarksnumbered=true,
	bookmarksopen=true,
	bookmarksopenlevel=1,
	linkcolor=blue,
	colorlinks=false,
	pdfstartview=Fit,
	pdfpagemode=UseOutlines,    
	pdfpagelayout=TwoPageRight
}

\DeclarePairedDelimiter\ket{\lvert}{\rangle}
\DeclarePairedDelimiterX\braket[2]{\langle}{\rangle}{#1\,\delimsize\vert\,\mathopen{}#2}
\newcommand{\unit}[1]{\ensuremath{\, \mathrm{#1}}}
\begin{document}

	\title{Optimisation of Active Space Debris Removal Missions With Multiple Targets Using Quantum Annealing}
	\author{Thomas Swain
	\vspace{0.5in}\\
	\href{mailto:thomas.swain@cantab.net}{thomas.swain@cantab.net}\\
	\href{mailto:thomas.swain@alumnos.upm.es}{thomas.swain@alumnos.upm.es}
	\vspace{0.5in}\\
		Under the direction of\\
		Alberto García\\
		Universidad Politécnica de Madrid\\
		\vspace{1in}
	}

	\maketitle

	\begin{abstract}
		A strategy for the analysis of active debris removal missions targeting multiple objects from a set of objects in near-circular orbit with similar inclination is presented. Algebraic techniques successfully reduce the orbital mechanics regarding specific inter-debris transfer and disposal methods to simple computations, which can be used as the coefficients of a quadratic unconstrained binary optimisation (QUBO) problem formulation which minimises the total propellant used in the mission whilst allowing for servicing time and meeting the mission deadline. The QUBO is validated by solving artificial small problems (from 2 to 11 debris) using classical computational methods and the weaknesses in using these methods are examined prior to solution using quantum annealing hardware. The quantum processing unit (QPU) and quantum-classical hybrid solvers provided by D-Wave are then used to solve the same small problems, with attention paid to evident strengths and weaknesses of each approach. Hybrid solvers are found to be significantly more effective at solving larger problems. Finally, the hybrid method is used to solve a large problem using a real dataset. From a set of 79 debris objects resulting from the destruction of the Kosmos-1408 satellite, an active debris removal mission starting on 30 September 2023 targeting 5 debris objects for disposal within a year with 20 days servicing time per object is successfully planned. This plan calculates the total propellant cost of transfer and disposal to be $ 0.87 \unit{km} / \unit{s} $ and would be complete well within the deadline at 241 days from the start date. This problem uses 6,478 binary variables in total and is solved using around $ 25 \unit{s} $ of QPU access time.
	\end{abstract}

	\tableofcontents
	\listoffigures
	\listoftables

	\chapter{Introduction}\label{intro}

	\section{Context}\label{context}

	Since the launch of satellite ``Sputnik 1" in 1957 by the Soviet Union, space agencies and private enterprises have continued to explore and develop near-Earth space through rockets, satellites, and other kinds of spacecraft. This exploration has created a significant amount of space debris and future space operations are potentially at risk due to the large number of uncontrolled objects occupying low Earth orbit (LEO) and geostationary orbit (GEO). A report by the European Space Agency (ESA) estimates that there are 29,000 objects larger than $ 10 \unit{cm} $, 670,000 objects larger than $ 1 \unit{cm} $ and more than 170,000,000 objects larger than $ 1 \unit{mm} $ in near-Earth orbit\cite{esa_num_debris}. Even objects with a size of $ 1 \unit{mm} $ have the potential to destroy sub-systems on board a spacecraft, and collision with a $ 10 \unit{cm} $ object would cause catastrophic fragmentation of a typical satellite. Astronauts are directly at risk, for example the International Space Station experienced a near miss in 2021, when mission managers discovered a small hole in a robotic arm caused by impact with space debris\cite{iss_near_miss}.

\begin{figure}[ht]
		\centering
		\includegraphics[width=0.8\textwidth]{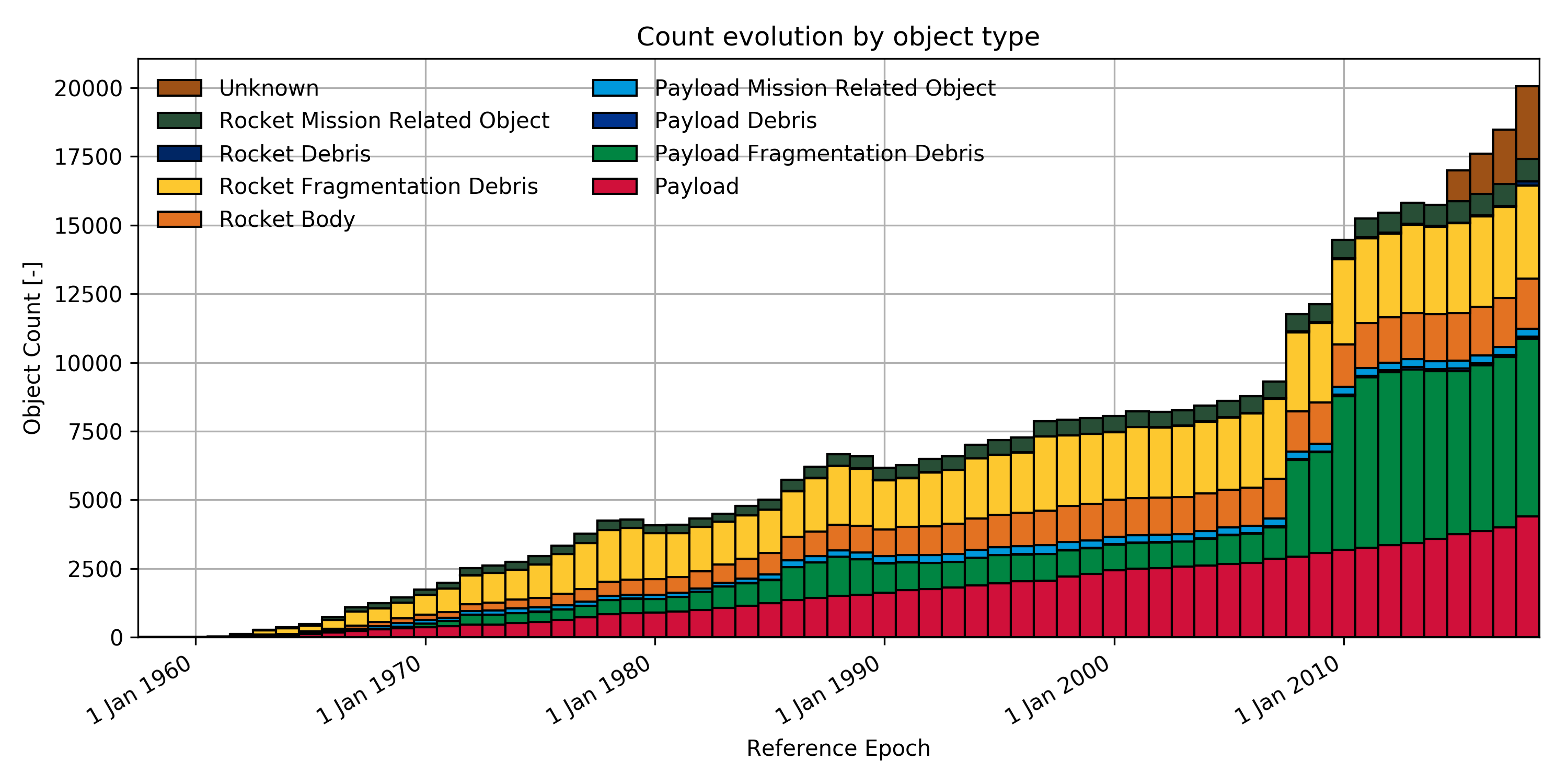}
		\caption{Number of various types of orbital debris since 1960}
		\label{esa_debris_count}
	\end{figure}

	Of particular concern is the Kessler syndrome, which states that each additional collision between debris can result in even more space debris being created which further increases the risk of collision, causing an exponential increase in the number of objects in orbit\cite{kessler}. This theoretical idea aside, a count provided by ESA shown in Figure \ref{esa_debris_count} empirically shows the increasing number of these undesired and uncontrolled objects in space since 1960\cite{debris_count_graph}.

\begin{figure}[ht]
		\centering
		\includegraphics[width=0.8\textwidth]{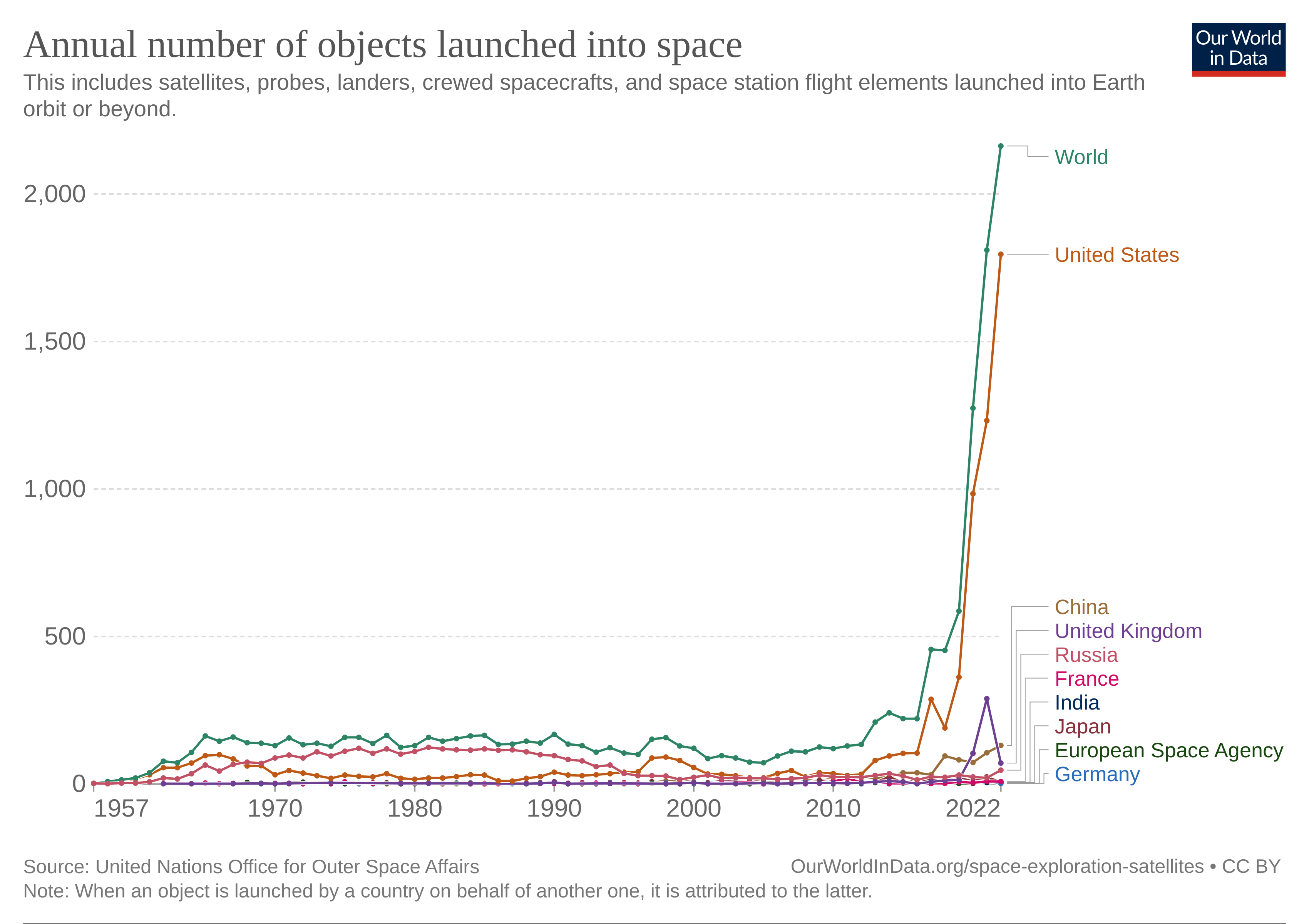}
		\caption{Annual number of objects launched into space}
		\label{unoosa_launch_count}
	\end{figure}

	In parallel the number of new objects launched has increased very rapidly over recent years and this trend is forecast to continue in the near future. For example, SpaceX has launched more than 4,700 \textit{Starlink} satellites into orbit already, has approval to deploy around 12,000 more and has applied for permission to launch another 30,000\cite{spacex}. The startling increase in the rate at which objects have been launched into space recently can be seen in Figure \ref{unoosa_launch_count} which shows a count of objects (including satellites, probes, landers and other types of objects) launched into space over past decades, based on data collected by the United Nations Office for Outer Space Affairs (UNOOSA)\cite{unoosa_launch_count}. This increase in space activity brings the risk associated with space debris to an all-time high.

	In response to these growing concerns, frameworks such as Space Surveillance and Tracking (SST) have emerged, which aim to reduce the risk of collisions in space by tracking debris objects using radars, telescopes and other means\cite{esa_sst}. Also, the Inter-Agency Space Debris Coordination Committee (IADC) has been formed to facilitate international collaboration between space agencies with an aim to tackle the problem of space debris and has published mitigation guidelines outlining steps that should be taken in the design of space operations to minimise debris creation\cite{iadc_guidelines}\cite{iadc}.

	In addition to the careful design of future space missions to minimise space debris left in the aftermath, space debris can be actively disposed of in dedicated missions. This is known as active debris removal (ADR), and there exist several methods to perform it. For example, Astroscale's \textit{COSMIC} mission and ESA's \textit{ADRIOS} mission plan to use a spacecraft to capture debris objects with a robotic arm and release them into a lower disposal orbit where the debris can burn up in the atmosphere\cite{ESAESAco72:online}\cite{Astrosca94:online}. Also, propulsive decommisioning kits installed before launch such as those manufactured by D-Orbit have the potential to be adapted for on-orbit installation on objects already in space during active debris removal missions\cite{ESAAward5:online}\cite{d-orbit}. It has been shown that at least five objects per year should be actively removed from near-Earth orbit to avoid the exponential increase in debris population predicted by the Kessler syndrome\cite{sensitivity_study}\cite{controlling_growth}.

	Whichever specific method is used to perform the disposal in ADR missions of this kind, key operational parameters defining the mission involve the total time needed for completion and the total propellant required by the spacecraft and in order to be commercially viable such an ADR mission must be planned effectively and seek optimal or at least satisfactory values for parameters like these. The task of planning effectively equates to formulation and solution of an optimisation problem, and a variety of computational methods and orbital transfer models have been studied and applied to this problem over recent years.

	This work focuses on ADR missions that remove multiple debris objects using a single spacecraft, and previous work with a similar focus that guides and motivates this work is summarised here.

	A plan for an ADR mission was obtained by using a branch and bound algorithm applied to a formulation comprised of a functional optimisation to define the orbital maneuvers and a combinatorial component stemming from the well-known travelling salesman problem (TSP) archetype to select the debris\cite{eads_adr}.

	Similar approaches using the time-dependent traveling salesman problem (TDTSP) formulation separately solved with timeline club optimisation (TCO)\cite{tdtsp_adr} and a combination of a deep neural network (DNN) plus reinforcement learning (RL)\cite{ml_adr} were also proposed.

	A TSP formulation using integer programming along with reasonable approximations of the orbital transfer propellant cost is of particular note and motivates a significant part of this work\cite{shen_lorenzo}.

	The Fujitsu quantum-inspired digital annealer was used alongside an artifical neural network and achieved results much faster than a domain expert working on the same problem using prior methods\cite{fujitsu_adr}.

	Last year, the application of quantum computing (QC) to solve the space debris problem was discussed with optimism\cite{artificial_brain_wp} and the D-Wave Systems quantum annealer has already proven to be a formidable device in this regard. For example, results from the quantum annealer were used to seed genetic algorithms in order to solve the electric vehicle charger placement problem, which significantly improved the results obtained\cite{evcp_qa}.

	The historical success of description of the space debris removal problem using TSP-like formulations and their optimisation using quantum annealing provides the motivation for this work and it is further developed throughout. This work is centered around optimisation of active debris removal missions with multiple targets using quantum annealing (QA).

	\section{Objectives}\label{objectives}

	The objectives of this work are as follows:

	\begin{itemize}
		\item[-] Construct a mathematical formulation to reflect a clearly defined ADR mission strategy
		\item[-] Analyse the complexity of the formulation
		\item[-] Test, examine and validate the formulation by solving artificial problems using classical methods
		\item[-] Test, examine and validate the formulation by solving artificial problems using quantum and hybrid methods
		\item[-] Use the strategy to plan an ADR mission to dispose of a set number of debris from the Kosmos-1408 cluster
	\end{itemize}

	\section{Outline}\label{outline}

	The work presented here is organised as follows:

	The remainder of Chapter \ref{intro} summarises the real-world problem this work aims to solve, outlines the type of optimisation problems of interest and describes how quantum annealing is well-suited to solve them.

	Chapter \ref{technical_formulation} derives and explains the exact mathematical model used to solve the problem.

	Chapter \ref{implementation} outlines the classical and quantum hardware and algorithms used to validate the model.

	Chapter \ref{analysis} describes the results obtained during validation of the model.

	Chapter \ref{practical_application} demonstrates the use of the model in a real-world practical scenario.

	Chapter \ref{conclusion} concludes this work.

	\section{Business Problem}\label{business_problem}

	This work considers an ADR mission where a single spacecraft transfers between a set of debris objects using propulsive maneuvers and performs disposal of each debris object via servicing that takes place purely on-orbit. The important mission-defining considerations are the propellant required for the spacecraft to dispose of and transfer between the target debris and the maximum time allowed for the mission.

	The specific question this work aims to answer is:

	\begin{quote}
	Which debris objects should the spacecraft dispose of, and in what order, such that the target number of debris is successfully disposed of before the mission deadline and the total propellant used by the spacecraft is minimised?
	\end{quote}

	This question can be answered by phrasing it mathematically as a combinatorial optimisation problem and using computers to perform the optimisation. The remainder of this chapter introduces the general mathematical and computational approach used in this work to solve the business problem outlined here.

	\section{Quadratic Unconstrained Binary Optimisation}\label{qubo}

	An important and well-studied subfield of mathematical optimisation is combinatorial optimsation, that presents the challenge of finding an optimal combination of objects from a finite set \cite{enwiki:1147380347}. Some examples of this type of problem are the travelling salesman problem (TSP) where the route to visit a group of cities must be chosen such that the total distance travelled is minimised, and the knapsack problem where the contents of a knapsack must be chosen such that the total value of the contents is maximised but the total weight of the contents does not exceed the knapsack's limits.

	In many cases, combinatorial optimisation problems can be formulated as quadratic unconstrained binary optimisation (QUBO) problems. This type of problem is a core concept throughout this work so it is outlined here.

	A QUBO introduces the set of binary vectors of length $ n > 0 $ representing the space of possible combinations of objects chosen from the set, denoted by $ \mathbb{B}^n $ where $ \mathbb{B} = \{0,1\} $. The number of possible binary vectors is $ |\mathbb{B}^n| = 2^n $, growing exponentially with the size $ n $ of the object set under consideration.

	A binary vector is referred to as $ \textbf{x} $, with each of its constituent binary variables referred to as $ x_i $ where $ 1 \le i \le n $. A real-valued upper triangular matrix $ \textbf{Q} \in \mathbb{R}^{n\times n} $ is introduced whose entries $ Q_{ij} $ define a weight for each pair of binary variable indices $ i,j \in \{1,...,n\} $.

	The function $ f_Q : \mathbb{B}^n \rightarrow \mathbb{R} $ is defined, mapping each binary vector $ x $ to a real value in the following way:

	\begin{equation} \label{qubo_matrix_equation}
	f_Q(\textbf{x}) = \textbf{x}^T\textbf{Q}\textbf{x} = \sum_{i=1}^n\sum_{j=i}^nQ_{ij}x_ix_j =\sum_{i=1}^nQ_{ii}x_i + \sum_{i=1}^n\sum_{j=i+1}^nQ_{ij}x_ix_j
	\end{equation}

	It should be noted that the distinction between elements $ Q_{ii} $ (diagonal) and $ Q_{ij} $ (off-diagonal) elements of $ \textbf{Q} $ allows linear and quadratic terms with respect to $ x_i $ to be separated as in the final expression of equation \ref{qubo_matrix_equation}, owing to the fact that if $ x_i \in \{0, 1\} $ then $ x_i^2 = x_i $.

	The QUBO problem consists of finding a binary vector $ \textbf{x}^* $ that minimises the value of function $ f_Q $:

	\begin{equation} \label{qubo_arg_min}
	\textbf{x}^* = \underset{\textbf{x}\in\mathbb{B}^n}{arg\,min}\;f_Q(\textbf{x})
	\end{equation}

	Generally, $ \textbf{x}^* $ is not unique, meaning there may be multiple vectors that minimise the function $ f_Q $ to the same value.

	Though QUBO problems can be expressed in a simple and elegant way, they belong to a class of problems known to be NP-hard meaning that even commercial solvers such as \textit{IBM ILOG CPLEX Optimiser}\cite{CPLEXSma36:online} and \textit{Gurobi Optimiser}\cite{TheLeade41:online} designed to find optimal solutions to such problems can still struggle with realistic-sized problems, where computations can run for weeks without producing a high quality solution\cite{glover2019tutorial}. The next section introduces a modern computational paradigm that can offer improved results in solving these types of optimisation problems.

	\section{Adiabatic Quantum Computing and Quantum Annealing}\label{aqc_qa}

	Quantum computing (QC) is a rapidly growing research field, offering new means to solve complex computational challenges like QUBO problems as a result of quantum physics phenomena such as superposition and entanglement. The two leading paradigms used to solve problems using QC are gate-based quantum computing and adiabatic quantum computation (AQC).

	The gate-based model involves application of a sequence of unitary gates to quantum bits (qubits) and measurement of the qubits at the end of the computation.

	In contrast, AQC makes use of the adiabatic theorem which states:

	\begin{quote}
		A physical system remains in its instantaneous eigenstate if a given perturbation is acting on it slowly
	enough and if there is a gap between the eigenvalue and the rest of the Hamiltonian’s spectrum\cite{born}.
	\end{quote}

	In other words, if a quantum system can be prepared in the ground state of an initial Hamiltonian then perturbed slowly enough, the final state of the system should be the ground state of the final Hamiltonian after the system evolution. The final Hamiltonian can be carefully designed such that the final ground state encodes the solution to a mathematical problem we wish to solve. This allows AQC to be applied to QUBO problems very effectively, though it should be noted that AQC is a universal model for computation not centered around optimisation\cite{Albash_2018}. It has been shown to be polynomially equivalent to gate-based quantum computing\cite{aqc_eq_gate_based}.

	Quantum annealing (QA) is very closely related to AQC, the distinction between them being subtle and related to history\cite{Albash_2018}, so for the purposes of this work, QA and AQC are treated as equivalent.

	The quantum annealing hardware provided by D-Wave Systems can be readily used remotely to solve problems correctly formulated as QUBO. The optimisation of active debris removal missions with multiple targets is formulated this way in the next chapter.

	\chapter{Technical Formulation}\label{technical_formulation}

	\section{Orbital Mechanics}\label{orbital_mechanics}

	Analysis of the orbital mechanics of debris in low Earth orbit allows mathematical approximations to be made, which motivate the specific formulation of the debris disposal problem outlined throughout this chapter.

	The motion of objects in orbit can be described using six osculating orbital elements (semimajor axis $ a $, eccentricity $ e $, inclination $ i $, right ascension of the ascending node (RAAN) $ \Omega $, argument of periapsis $ \omega $ and mean anomaly $ M $) at some reference time, combined with a propagation model that computes properties in the future relative to this time. The propagation model is necessary because the orbital properties vary as a result of perturbations such as Earth oblateness, relativistic effects, atmospheric drag and radiation pressure, to name a few.

	A common representation of this reference data is in two-line element (TLE) sets which are available freely from CelesTrak and updated daily\cite{celestrak}. A well-known propagation model is SGP4 which takes into account many sources of orbit perturbation, however in this work only perturbations due to Earth oblateness (J2) are considered to simplify formulation of the problem.

	Given the osculating elements of a debris object at a specifc reference time obtained from a single TLE, the semimajor axis $ a $, eccentricity $ e $ and inclination $ i $ are treated as constants and can be used to calculate the steady rate of change of RAAN $ \Omega $ according to\cite{shen_lorenzo}

	\begin{equation} \label{d_raan_d_t}
	\dot{\Omega} = \frac{d\Omega}{dt} = -\frac{3}{2}\left(\frac{r_E}{p}\right)^2nJ_2\cos{i}
	\end{equation}

	where $ n = \sqrt{\mu/a^3} $ is the mean motion, $ p = a(1 - e^2) $ is the semilatus rectum, $ r_E = 6378,000 \unit{m} $ is the radius of the Earth at the equator, $ \mu = 3.986004418 \times 10^{14} \unit{m}^3 / \unit{s}^2 $ is Earth's gravitational parameter and $ J_2 = -0.1082635854 \times 10^{-2} $ is the second zonal harmonic of the geopotential model. Propagation approximation using only this rate of change of $ \Omega $ provides great accordance with SGP4 for the orbital elements required in this work, even for 200 days after the reference time\cite{shen_lorenzo}.

	Using this rate, the time that two different debris objects $ j $ and $ k $ have their $ \Omega $ aligned after the reference time $ t_0 $ is given by

	\begin{equation} \label{T_ij}
	T_{j,k} = \frac{\Omega_j(t_0) - \Omega_k(t_0) + K.2\pi}{\dot{\Omega}_k - \dot{\Omega}_j}
	\end{equation}

	where the arbitrary integer $ K $ is selected such that $ T_{j,k} $ sensibly refers to the first alignment after $ t_0 $. The appropriate value for $ K $ is therefore set explicitly depending on $ \Omega_j(t_0) $, $ \Omega_k(t_0) $, $ \dot{\Omega}_j $ and $ \dot{\Omega}_k $, giving an easy way to compute this for every pair of debris under consideration:

	\begin{equation}\label{t_ij_explicit}
		K =
		\begin{cases}
			0 & \text{if $ \Omega_j(t_0) - \Omega_k(t_0) \ge 0 $ and $ \dot{\Omega}_k - \dot{\Omega}_j \ge 0 $} \\
			0 & \text{if $ \Omega_j(t_0) - \Omega_k(t_0) < 0 $ and $ \dot{\Omega}_k - \dot{\Omega}_j < 0 $} \\
			-1 & \text{if $ \Omega_j(t_0) - \Omega_k(t_0) \ge 0 $ and $ \dot{\Omega}_k - \dot{\Omega}_j < 0 $} \\
			1 & \text{if $ \Omega_j(t_0) - \Omega_k(t_0) < 0 $ and $ \dot{\Omega}_k - \dot{\Omega}_j \ge 0 $} \\
		\end{cases}
	\end{equation}

	At time $ T_{j,k} $, the RAAN of the orbits of debris $ j $ and $ k $ are aligned, so transfer by the spacecraft between them can be obtained with a small amount of propellant consumption using the Hohmann transfer\cite{shen_lorenzo}. The propellant cost required to transfer between orbits using this method is approximated by\cite{delta_v}

	\begin{equation} \label{C_ij}
	C_{j,k} = \Delta V_{j,k} = \frac{1}{2}V_j\sqrt{(\Delta a / a_j)^2 + (\Delta e)^2 + (\Delta i)^2}
	\end{equation}

	where $ j $ is assumed by convention to denote the orbit with the smallest semimajor axis $ a $. The velocity $ V_j $ is calculated assuming a circular orbit by $ V_j = \sqrt{\mu / a_j} $. Since the ensembles of debris targeted during this work have small eccentricity, and the proportional differences in semimajor axis, eccentricity and inclination between the debris objects are small, this approximation of transfer cost is sufficient for the formulation proposed in this work.

	It should be noted that both $ T_{j,k} $ and $ C_{j,k} $ defined in this section are symmetrical with respect to $ j $ and $ k $.

	Using Edelbaum's approximation\cite{edelbaum}, again owing to the small eccentricities and therefore near-circular orbits of the debris considered, the propellant cost required to dispose of the debris via uncontrolled reentry is given by

	\begin{equation} \label{C_i}
	C_{j} = \Delta V_{j} = \sqrt{(\mu / r_p)} - \sqrt{(\mu / a_j)}
	\end{equation}

	where $ r_p = 1.02r_E $, corresponding to an altitude of about 125 km above Earth's surface.

	The quantities $ T_{i,j } $, $ C_{i,j} $ and $ C_i $ for a given debris $ i $ or debris pair $ i,j $ defined in this section can be expressed using matrices and a vector. For example with a total of 2 debris, the following equations for $ \textbf{T} $, $ \textbf{C} $ and $ \textbf{c} $ succinctly describe any problem instance:

	\begin{equation} \label{t_ij_2}
	\textbf{T} = \begin{pmatrix}
					 T_{1,1} & T_{1,2}\\
					 T_{2,1} & T_{2,2}\\
	\end{pmatrix}
	\end{equation}

	\begin{equation} \label{c_ij_2}
	\textbf{C} = \begin{pmatrix}
					 C_{1,1} & C_{1,2}\\
					 C_{2,1} & C_{2,2}\\
	\end{pmatrix}
	\end{equation}

	\begin{equation} \label{c_i_2}
	\textbf{c} = \begin{pmatrix}
					 C_{1}\\
					 C_{2}\\
	\end{pmatrix}
	\end{equation}

	As a final note about this section, it should be emphasised that these problem-defining variables are indeed calculated using approximations and any solutions found based on the analysis here should be subject to further calculations and checks to verify that propagation estimates are sufficiently accurate.

	The analysis done here aims to reduce the very complicated subject matter of debris orbital mechanics to simple computations that can feed the QUBO formulation effectively at the expense of absolute accuracy. The exact transfer and disposal strategies can be refined further using accurate calculations once an ordered set of debris has hopefully been found using the computational methods proposed in this work.

	\section{Formulation}\label{core_formulation}

	The optimisation problem is formulated in a graphical manner similar to the traveling salesman problem, where each piece of debris is considered as a node.

	Movement from any debris $ i $ to any other $ j $ is considered an edge, denoted by the binary variable $ x_{i,j} $ defined by:

	\begin{equation}
		x_{i,j}=
		\begin{cases}
			1 & \text{if debris $ i $ transfers to debris $ j $} \\
			0 & \text{otherwise}
		\end{cases}
	\end{equation}

	Only binary variables with $ i \neq j $ are considered as it is nonsensical that debris $ i $ would transfer to itself.

	The total number of real debris is $ N_t $ and a dummy piece of debris numbered $ 0 $ is introduced to help add constraints to the formulation.

	The cost of transferring from debris $ i $ to debris $ j $ is described by $ C_{i,j} $ with $ C_{0,i} = C_{i,0} = 0 \;\; \forall i : i > 0 $ and the cost of disposal of debris $ i $ is described by $ C_{i} $ with $ C_{0} = 0 $, therefore the main objective function to be minimized is:

	\begin{equation} \label{minimise_cost}
	H = \sum_{i=0}^{N_t}\sum_{\substack{j=0\\j\neq i}}^{N_t}x_{i,j}(C_{i,j} + C_{i})
	\end{equation}

	It should be noted that a convention is chosen that includes the disposal cost of debris $ i $ only if there is a transfer from debris $ i $ (in other words, it has a departure). This convention is made effective by constraint 4, described in the following.

	The first constraint forces the total number of debris selected to be equal to a fixed value $ N_s $ ($ N_s + 1 $ including dummy debris 0):

	\begin{equation} \label{constraint_1}
	\sum_{i=0}^{N_s}\sum_{\substack{j=0\\j\neq i}}^{N_s}x_{i,j} = N_s + 1
	\end{equation}

	With some algebraic manipulation, a function $ C_1 $ representing this constraint is obtained that is equal to zero if the constraint is satisfied and greater than zero if not. Minimizing this function so its value is equal to 0 is therefore equivalent to satisfying the constraint:

	\begin{equation} \label{constraint_1_qubo}
	C_1 = \left(\sum_{i=0}^{N_s}\sum_{\substack{j=0\\j\neq i}}^{N_s}x_{i,j} - (N_s + 1)\right) ^ 2
	\end{equation}

	The second constraint says that the dummy debris 0 must have exactly 1 departure:

	\begin{equation} \label{constraint_2}
	\sum_{i=1}^{N_t}x_{0,i} = 1
	\end{equation}

	\begin{equation} \label{constraint_2_qubo}
	C_2 = \left(\sum_{i=1}^{N_t}x_{0,i} - 1\right)^2
	\end{equation}

	Similarly, the third constraint says that the dummy debris 0 must have exactly 1 arrival:

	\begin{equation} \label{constraint_3}
	\sum_{i=1}^{N_t}x_{i,0} = 1
	\end{equation}

	\begin{equation} \label{constraint_3_qubo}
	C_3 = \left(\sum_{i=1}^{N_t}x_{i,0} - 1\right)^2
	\end{equation}

	The fourth constraint says that all the real debris have 0 or 1 departures:

	\begin{equation} \label{constraint_4}
	\sum_{\substack{j=0\\j \neq i}}^{N_t}x_{i,j} \le 1 \;\;\;\;\forall i : i > 0
	\end{equation}

	Since this constraint is an inequality, a binary slack variable $ s_{4,i} $ is introduced for any given $ i $, and the constraint is converted into a function that is only equal to zero if satisfied for all $ i $:

	\begin{equation} \label{constraint_4_qubo}
	C_4 = \sum_{i=1}^{N_t}\left(\sum_{\substack{j=0\\j \neq i}}^{N_t}x_{i,j} + s_{4,i} -  1\right)^2
	\end{equation}

	Similarly, constraint 5 says that all the real debris have 0 or 1 arrivals:

	\begin{equation} \label{constraint_5}
	\sum_{\substack{j=0\\j \neq i}}^{N_t}x_{j,i} \le 1 \;\;\;\;\forall i : i > 0
	\end{equation}

	\begin{equation} \label{constraint_5_qubo}
	C_5 = \sum_{i=1}^{N_t}\left(\sum_{\substack{j=0\\j \neq i}}^{N_t}x_{j,i} + s_{5,i} -  1\right)^2
	\end{equation}

	Constraint 6 forces the number of departures from a node to be equal to the number of arrivals to it.

	\begin{equation} \label{constraint_6_and_7_better}
	\left( \sum_{\substack{i=0\\i\ne j}}^{N_t}x_{i,j}-\sum_{\substack{k=0\\k\ne j}}^{N_t}x_{j,k}\right )^2 = 0\;\;\;\;\forall j : j > 0
	\end{equation}

	\begin{equation} \label{constraint_6_and_7_qubo_better}
	C_6 = \sum_{j=1}^{N_t}\left( \sum_{\substack{i=0\\i\ne j}}^{N_t}x_{i,j}-\sum_{\substack{k=0\\k\ne j}}^{N_t}x_{j,k}\right )^2
	\end{equation}

	Constraint 7 forces the departure away from a given node to not return to the node visited previously:

	\begin{equation} \label{constraint_6_and_7_qubo_better}
	C_7 = \sum_{i=0}^{N_t}\sum_{j=i+1}^{N_t}x_{i,j}x_{j,i}
	\end{equation}

	Constraint 8 is concerned with timing. If the spacecraft arrives at debris $ j $ from debris $ i $ at time $ T_{i,j} $, the time at which the spacecraft departs debris $ j $ and travels to debris $ k $ cannot be before $ T_{i,j} + T_s $, to allow for on-orbit servicing of debris $ j $. Therefore, if $ T_{i,j} + T_s > T_{j,k} $ then it must be that $ x_{i,j}x_{j,k} = 0 $:

	\begin{equation} \label{constraint_8_efficient}
	x_{i,j}x_{j,k} = 0 \;\;\;\;\forall i,j,k : j > 0,\;i \neq j,\;k \neq j,\;k \neq i,\;T_{i,j}+T_s>T_{j,k}
	\end{equation}

	\begin{equation} \label{constraint_8_efficient_qubo}
	C_8 =  \sum_{i = 0}^{N_t}\sum_{\substack{j=1\\j \neq i}}^{N_t}\sum_{\substack{k=0\\k \neq i\\k \neq j\\T_{i,j}+T_s>T_{j,k}}}^{N_t}x_{i,j}x_{j,k}
	\end{equation}

	Regarding the dummy node $ 0 $, $ T_{0,i} = 0 \; \forall i : i > 0 $, allowing node 0 to be the first departure. Also, the mission has a deadline $ T_{\max} $, therefore the arrival time at node 0 is set $ T_{i,0} = T_{\max} \; \forall i : i > 0 $ so that the final edge arriving at node 0 always honours the deadline.

	It should also be noted that in this constraint, $ j \neq 0 $ because node $ 0 $ as the intermediate node will always violate this constraint, but node $ 0 $ must be allowed to be an intermediate node.

	The final function to be minimised is simply a sum of the expression representing the total cost in terms of propellant and each constraint expression, with Lagrange multipliers $ L_x $ used to apply a weighting to each component:

	\begin{equation} \label{final_qubo_core}
	F = L_H H + \sum_{c=1}^8 L_c C_c
	\end{equation}

	\section{Complexity Analysis}\label{complexity_analysis}

	The complexity of the formulation with respect to the total number of debris considered $ N_t $ and number to select $ N_s $ is best illustrated with a fully-connected, directed graph with $ N_t + 1 $ nodes and $ N_t(N_t + 1) $ edges.

	The number of binary variables required by the formulation (not including slack variables) is equal to the number of edges. Inclusion of the slack variables adds an extra $ 2 N_t $ variables, bringing the total to $ N_t(N_t + 3) = O(N_t^2). $

	The total number of valid paths through $ N_s $ distinct nodes on the graph (starting and ending at node 0) is equal to $ \frac{N_t!}{(N_t - N_s)!} = O({N_t}^{N_s}) $. This idea of a valid path disregards constraint 8 and is raised here simply to demonstrate that the number of possible paths through a graph grows exponentially with the path length for a given total number of nodes.

\begin{figure}
		\centering
		\begin{tabular}{cc}
			\includegraphics[width=0.45\textwidth]{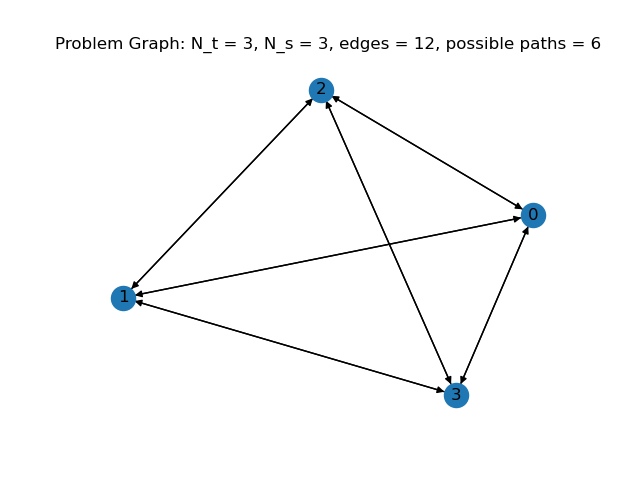} &   \includegraphics[width=0.45\textwidth]{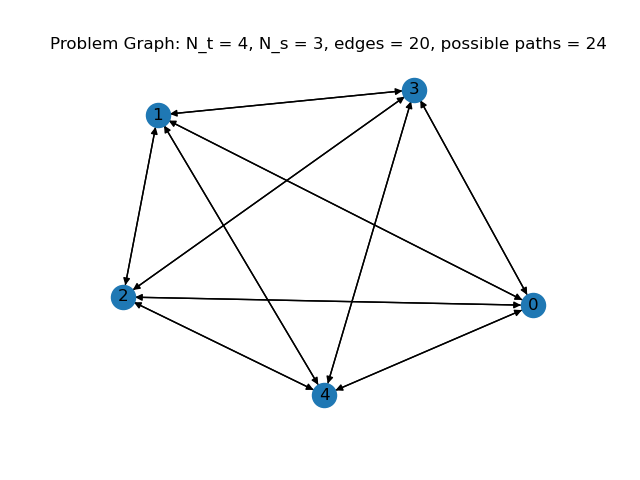} \\
			\includegraphics[width=0.45\textwidth]{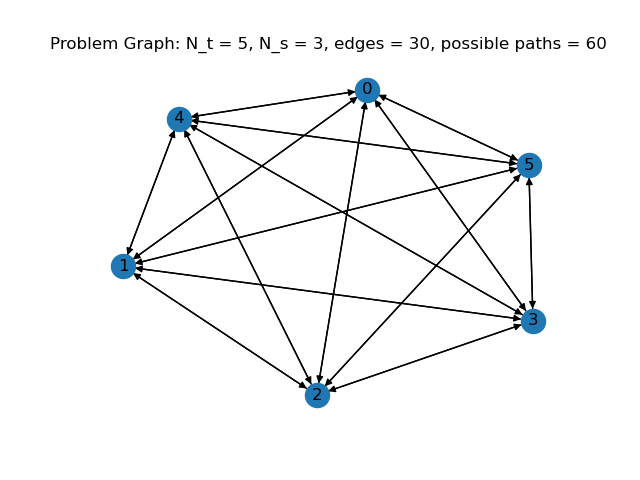} &   \includegraphics[width=0.45\textwidth]{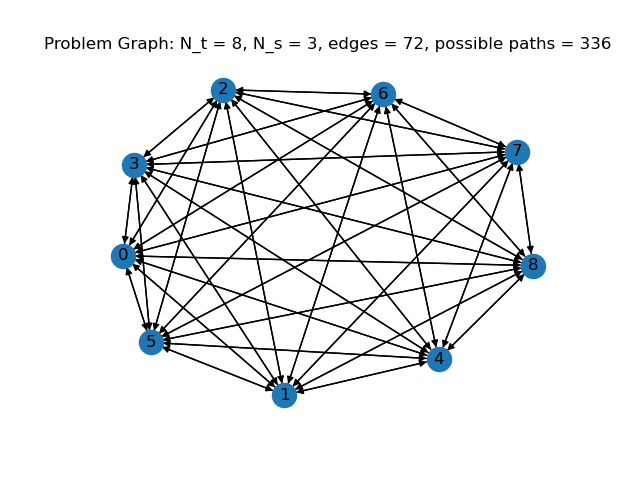} \\
			\includegraphics[width=0.45\textwidth]{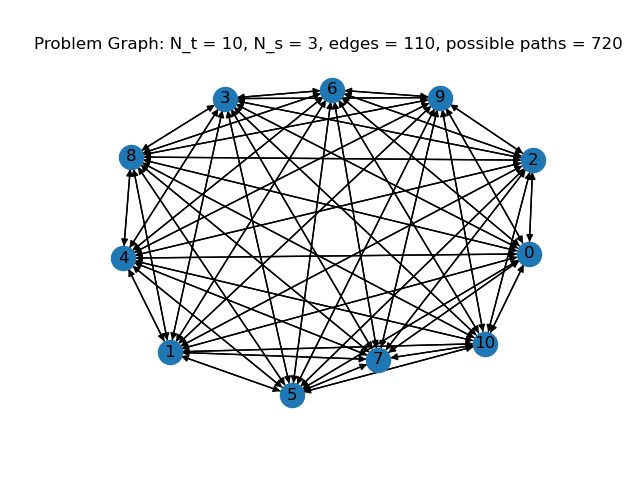} &   \includegraphics[width=0.45\textwidth]{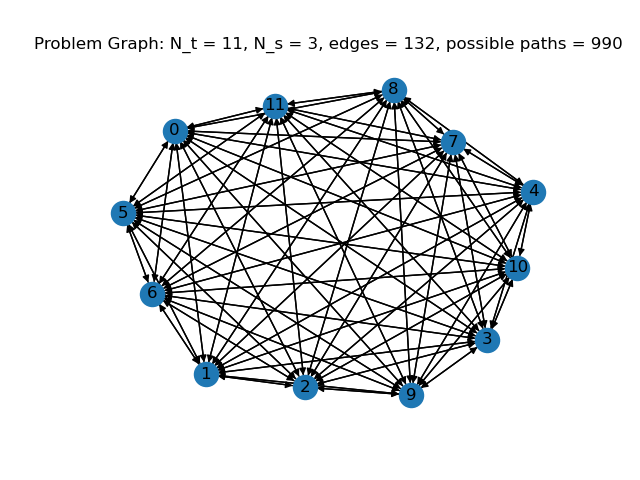} \\
		\end{tabular}
		\caption{Graphs illustrating problems of various sizes}
		\label{problem_graphs}
	\end{figure}

\begin{figure}[H]
		\centering
		\includegraphics[width=0.45\textwidth]{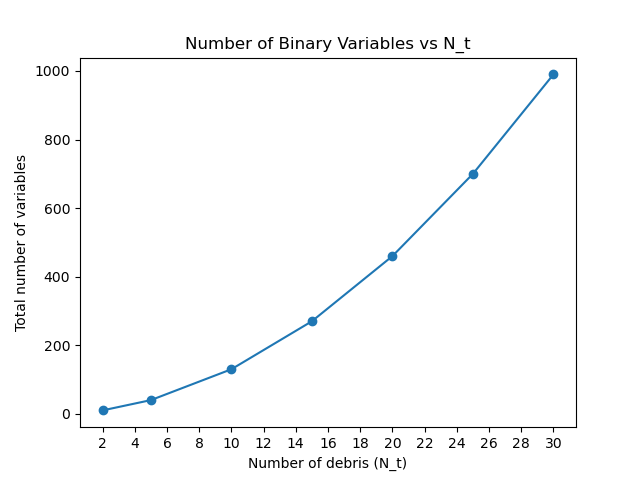}
		\caption{Binary variables with varying $ N_t $}
		\label{var_vs_deb}
	\end{figure}

	Growth of the size of the graph (in particular the number of edges or binary variables in the QUBO formulation) with $ N_t $ is demonstrated visually in Figure \ref{problem_graphs}. It should be noted that each edge represents two directed edges (the arrows are bi-directional). The same relationship is shown numerically in Figure \ref{var_vs_deb}. With 30 debris in total, the QUBO formulation requires a total of around 1,000 binary variables to express the problem.

\begin{figure}[H]
		\centering
		\includegraphics[width=0.45\textwidth]{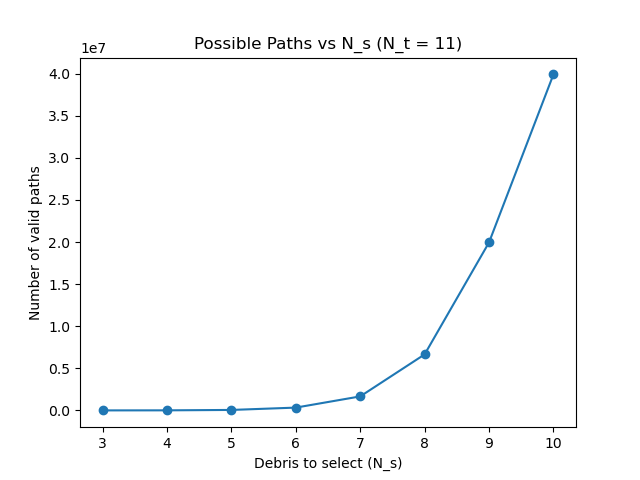}
		\caption{Valid paths with varying $ N_s $ on graphs with $ N_t = 11 $}
		\label{possible_paths}
	\end{figure}

	For $ N_t = 11 $, the exponential growth of the number of valid paths through the graph is shown in Figure \ref{possible_paths}. As can be seen, there are around 40,000,000 valid paths when selecting 10 debris from a set of 11.

	\chapter{Implementation}\label{implementation}

	To test the efficacy of this QUBO formulation for solving problems using quantum annealing hardware, a series of small-sized problems are initially solved using various classical algorithms to empirically see the effect of $ N_t $ and $ N_s $ on the solution accuracy, and to have a rough strategy for selecting the Lagrange multipliers of equation \ref{final_qubo_core}.

	Afterwards, the problem is executed both directly on a quantum annealer and using a hybrid algorithm which uses state-of-the-art classical optimisation algorithms together with quantum annealing to solve the problem.

	In all cases, \textit{Python} 3.9.16 is the programming language used for general programming and \textit{PyQUBO} 1.4.0 is used to create the QUBO models\cite{python_docs}\cite{pyqubo}. This chapter provides an outline of the algorithms and hardware used, prior to analysis of the execution results in the next chapter.

	Throughout this work, for solutions obtained by algorithms that are probabilistic and involve sampling, the accuracy is defined as the proportion of solution samples that are valid (satisfy all constraints) out of all samples. These types of algorithms should be executed a large number of times to provide solution samples. Depending on how well the formulation is constructed, how well the Lagrange multipliers are chosen and importantly the size and specific quantities defining the problem instance, the proportion of valid solutions in the group of samples is expected to vary significantly.

	\section{Classical Algorithms}\label{imp_classical}

	All classical algorithms were run directly on a local laptop computer with an Intel i7-8565U 1.80 GHz processor and 32 GB RAM, using the \textit{D-Wave Ocean} 6.0.1 suite of classical samplers\cite{dwave_samplers} which is freely available and executable on a local computer, so no access to or use of cloud resources was required.

	\subsection{Steepest Descent}\label{imp_steepest_descent}

	Steepest descent (SD), also called gradient descent, is a first-order iterative optimisation algorithm, effective at finding a local minimum\cite{sd_wiki}.

	From a starting point, the direction of the steepest downwards (negative) energy gradient is calculated and a step in that direction is taken to obtain the next point. This is repeated until no direction away from the current point has a negative energy gradient and the minimum has been reached.

	The obvious weakness of this type of search is that unless the starting point is a good initial guess and lies directly uphill from the global minimum, this algorithm is likely to get stuck in a local minimum if there are a significant number of them. Therefore this algorithm is suited to simple energy landscapes with no or a limited number of non-global minima, or to complicated energy landscapes supplemented with a good initial guess of the solution typically provided by the output of another algorithm.

	\subsection{Tabu Search}\label{imp_tabu_search}

	Tabu search (TS) is a metaheuristic search method that employs local search for mathematical optimisation\cite{ts_wiki}.

	Similarly to steepest descent, local search methods can easily become stuck in local minima. Tabu search enhances local search in this regard, by accepting moves that worsen the solution if no improving move is available, as is the case at the low of a local minimum.

	Furthermore, local minima are marked as prohibited (hence the name``Tabu") which discourages the algorithm from revisiting them. To keep track of the prohibitions, tabu search must maintain a list of them (the ``tabu list") in memory.

	This means tabu search can effectively find the global minimum of energy landscapes with many local minima, but the resources required to keep the tabu list grows directly with the number of them. The tabu search provided by D-Wave Ocean allows the tabu list to have maximum size of 20, and is known as the ``tenure" of the algorithm.

	\subsection{Simulated Annealing}\label{imp_simulated_annealer}

	Simulated annealing (SA) is a metaheuristic to approximate global optimisation in a large search space\cite{sa_wiki}.

	It is a probabilistic algorithm, and is able to find the global minimum of a problem even when there are large numbers of local minima. The SA algorithm takes inspiration from annealing in metallurgy, where heating and controlled cooling of a material can alter its physical properties.

	The central property of simulated annealing that makes it effective at finding global solutions is that worse solutions can be accepted during the algorithm's steps. At each time step, a solution close to the current one is randomly selected. If this solution is worse than the current one, the algorithm may still accept this solution for the next step, depending on the temperature. If this solution is better, the algorithm will definitely accept it. Higher temperatures mean the probability of accepting a worse solution is higher, and the temperature decreases throughout. 	As a result, the algorithm should explore the solution space thoroughly at the beginning of the algorithm and should settle close to the global minimum at the end.

	In the \textit{D-Wave Ocean} implementation, the probabilistic algorithm used to move between candidate solutions at each time step is a type of Markov chain Monte Carlo (MCMC) method called the Metropolis-Hastings algorithm\cite{metroplis_hastings}.

	\section{Quantum Annealing}\label{imp_quantum_annealing}

	Quantum annealing (QA), as outlined earlier in section \ref{aqc_qa}, works by designing the Hamiltonian of a quantum system such that the ground state encodes the solution to a problem.

	The D-Wave quantum processing unit (QPU) uses superconducting loops to implement the qubits of the quantum system in question, with the two lowest energy states of each loop representing the qubit states $ \ket{0} $ and $ \ket{1} $.

\begin{figure}[H]
		\centering
		\includegraphics[width=0.8\textwidth]{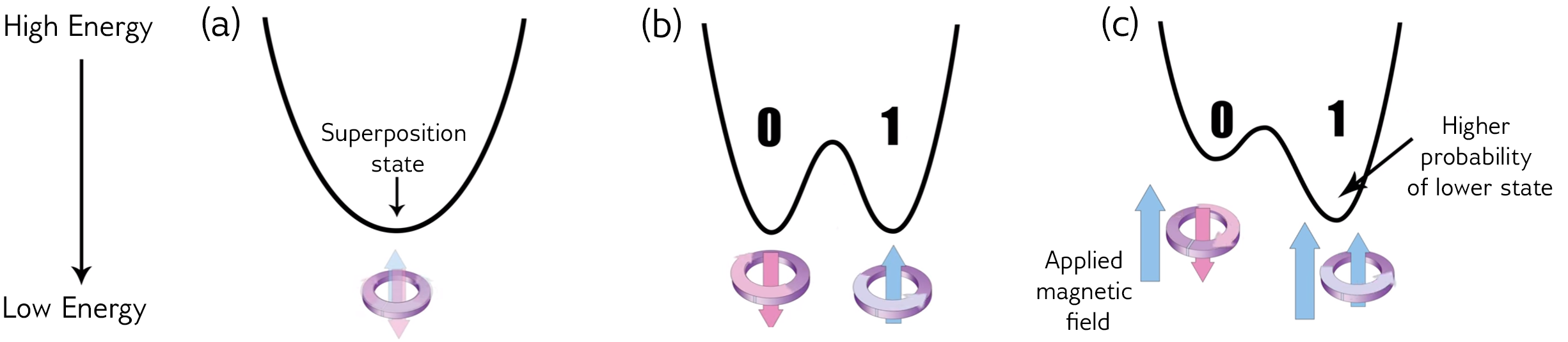}
		\caption{D-Wave quantum annealing operation}
		\label{dwave_qa}
	\end{figure}

	At the beginning of the anneal, each qubit is set such that there is only a single well so the qubit is in an equal superposition of states $ \ket{0} $ and $ \ket{1} $. The QPU then applies a magnetic field to each loop, gradually introducing problem-specific biases and coupling between the qubits, so the final quantum system has increased probability of representing the solution. This action is summarised in Figure \ref{dwave_qa} and a full description of the operation can be found in the D-Wave documentation\cite{dwave_qa}.

	The D-Wave \textit{Advantage} QPU has more than 5,000 qubits connected by the \textit{Pegasus} topology\cite{dwave_topology}, which directly limits which qubits of the hardware can be coupled with each other. With this topology, each qubit can be connected to 15 other qubits. Since the graphical problem studied in this work is fully connected (every binary variable has a quadratic term with every other), problems with 3 debris objects (corresponding to 18 variables in total) or greater requires chaining qubits within the QPU, whereby a single variable is mapped to multiple qubits with a purposefully strong coupling (quantified by a measure called ``chain strength") between them such that the measurement on each qubit is the same but the connectivity required to represent the problem is therefore possible.

	Throughout this work, one of \textit{Advantage} 4.1, 5.4 or 6.2 was used, which are all made available over the internet via \textit{D-Wave Leap}. When problems were submitted, the selection of a QPU from these versions was handled automatically by the backend and from this perspective they were treated as equivalent.

	\section{Quantum-classical Hybrid Algorithm}\label{imp_hybrid}

	Hybrid algorithms make use of quantum annealing but combine it with state-of-the-art classical computational methods, and \textit{D-Wave Leap} has a suite of such algorithms available\cite{dwave_hybrid}.

	An example of such algorithms include use of quantum annealing to find some solution states and subsequently seed algorithms like steepest descent with these states.

	From the user's perspective, it is enough to submit a problem formulated as a QUBO to the backend, and the exact hybrid algorithm used is automatically selected by \textit{D-Wave Leap} and the answer returned to the user if successful. Throughout this work, the \textit{Hybrid Binary Quadratic Model version 2} was used.

	\chapter{Results and Analysis}\label{analysis}

	To examine the performance of the algorithms, problem instances of various sizes were created artificially and are outlined in Appendix \ref{appendix_artificial_problems}.

\begin{figure}[H]
		\centering
		\includegraphics[width=0.55\textwidth]{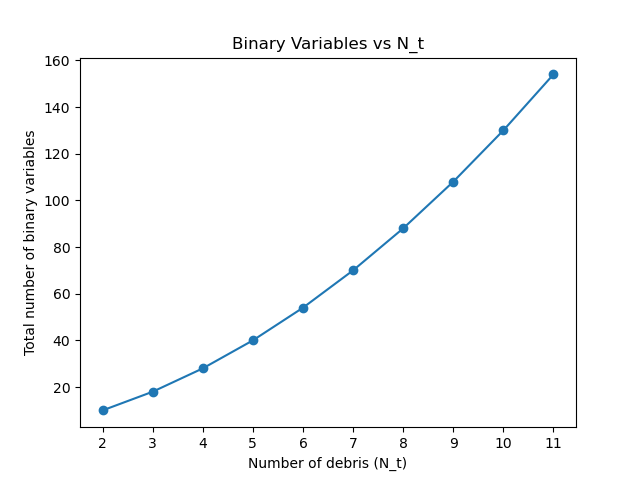}
		\caption{Binary variables required by artificially created problems}
		\label{var_vs_deb_small}
	\end{figure}

	The number of binary variables required to represent each artificial problem with total debris $ N_t $ is shown in Figure \ref{var_vs_deb_small} and is of particular relevance in this chapter. To summarise, the smallest problem considered with 2 debris objects requires 10 binary variables and the largest with 11 debris objects requires 154.

	The Lagrange multipliers must be chosen so that constraint satsifaction is strongly prioritised, so the strategy must certainly ensure that the multipliers for each constraint $ C_i $ are significantly larger than the multplier for the main cost function $ H $. This way, a solution that successfully minimises the main cost function by means of breaking constraints will have significantly higher energy in the complete Hamiltonian. This strategy combined with ordering of solution samples by total energy from smallest to largest will clearly identify the valid solutions offered by probabilistic algorithms.

	The following Lagrange multipliers were found to be effective in applying weight to the relevant components of the Hamiltonian, enforcing the constraints by ensuring the penalties are large enough:

	\begin{equation} \label{toy_lagrange}
	\begin{split}
		&L_H = 1, \;\;L_1 = 2500, \;\;L_2 = 300, \;\;L_3 = 300, \;\;L_4 = 300, \\
		&L_5 = 300, \;\;L_6 = 2500, \;\;L_7 = 4000, \;\;L_8 = 5000
	\end{split}
	\end{equation}

	These values were used to obtain the results described in this chapter. Strategically, $ L_H $ is set to $ 1 $ so that the energy of the Hamiltonian can be read directly as the total propellant cost associated with the solution.

	\section{Classical Algorithms}

	\subsection{Steepest Descent}\label{steepest_descent}

\begin{figure}[H]
		\centering
		\includegraphics[width=0.55\textwidth]{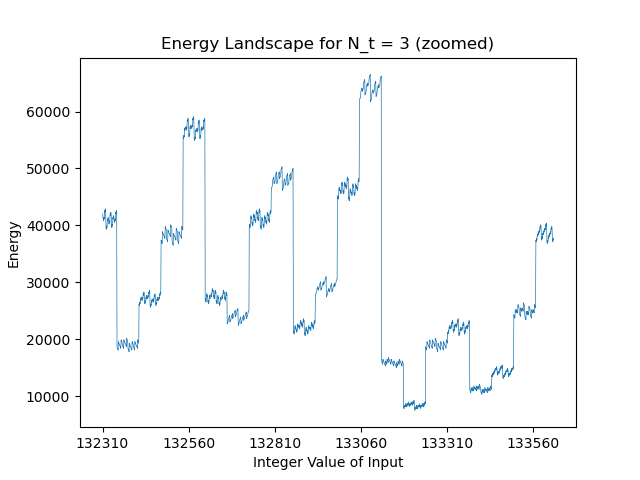}
		\caption{Small section of the energy landscape with $ N_t = 3 $ showing many local minima}
		\label{3_deb_energy_landscape}
	\end{figure}

	Since the minima of the energy landscape are directly relevant with regards to the steepest descent method, the energy landscape with $ N_t = 3 $ (18 binary variables) is examined in Figure \ref{3_deb_energy_landscape}. The x-axis shows a small subset (1 / 200) of all possible inputs to the binary variables defining the problem displayed as integers, and the y-axis shows the Hamiltonian energy as a result of the input.

	As can be seen, even with a tiny subset of the entire landscape for a very small-sized problem, there are multiple and regular minima. Therefore the risk of this algorithm obtaining an incorrect solution corresponding to the low of a non-global minimum is high, particularly in larger problems.

	To explicitly demonstrate this, the algorithm was seeded separately with two different initial states, the ``zero" state (all binary variables set to 0) and the state corresponding to the integer value 140035 which is 3 bit-flips away (so nearby and uphill) from the global minimum with integer value 140032.

	The algorithm ran to completion very quickly taking less than $ 2 \unit{ms} $ for both runs, but as expected the first run with the trivial initial state did not obtain the correct answer and returned a local minimum. The second run using the state close to the global minimum successfully obtained the correct solution, but given the large number of variables and local minima, finding the state with which to seed the algorithm almost equates to solving the original problem.

	\subsection{Tabu Search}\label{tabu_search}

\begin{figure}[H]
		\centering
		\includegraphics[width=0.55\textwidth]{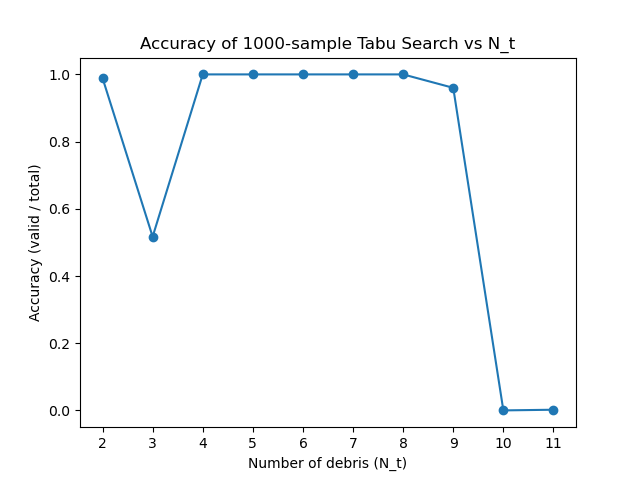}
		\caption{Tabu search accuracy with varying artificial problem instance size}
		\label{tabu_accuracy_plot}
	\end{figure}

\begin{figure}[H]
		\centering
		\includegraphics[width=0.55\textwidth]{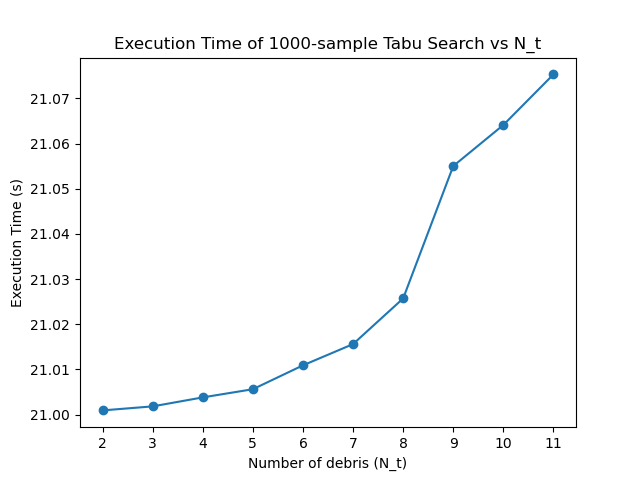}
		\caption{Tabu search execution time with varying artificial problem instance size}
		\label{tabu_time_plot}
	\end{figure}

	Tabu search was performed to obtain a total of 1,000 samples with varying values of $ N_t $ in the range $ 2 $ to $ 11 $. Plots of solution accuracy and execution time against $ N_t $ are shown in Figure \ref{tabu_accuracy_plot} and \ref{tabu_time_plot} respectively.

	As can be seen, the execution time is fairly constant at around 21 s with small additional time in the order of milliseconds over this range of $ N_t $. Nonetheless the execution time clearly increases with problem size, so it could take lengthy amounts of time to use this method to solve large problems.

	Also, it can be seen that the accuracy sharply reduces from near 100 \ to near 0 \ in changing $ N_t $ from 9 to 10 (number of binary variables changing from 108 to 130). This is possibly due to the maximum length of the tabu list being reached when the number of variables and energy landscape complexity increases, so the algorithm cannot consider all local minima effectively before completion.

	\subsection{Simulated Annealing}\label{simulated_annealer}

\begin{figure}[H]
		\centering
		\includegraphics[width=0.55\textwidth]{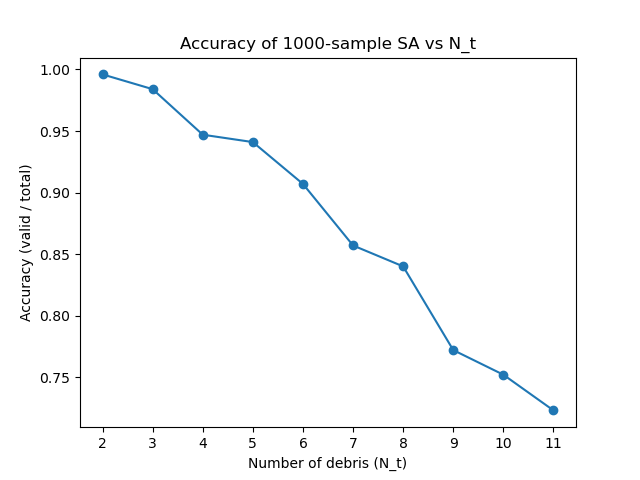}
		\caption{Simulated annealing accuracy with varying artificial problem instance size}
		\label{sa_accuracy_plot}
	\end{figure}

\begin{figure}[H]
		\centering
		\includegraphics[width=0.55\textwidth]{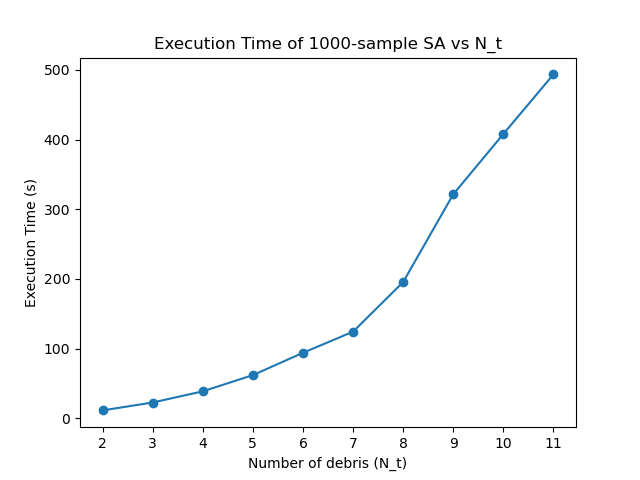}
		\caption{Simulated annealing execution time with varying artificial problem instance size}
		\label{sa_time_plot}
	\end{figure}

	Simulated annealing was performed to obtain a total of 1,000 samples of 50,000 sweeps with varying values of $ N_t $ in the range $ 2 $ to $ 11 $. Plots of solution accuracy and execution time against $ N_t $ are shown in Figure \ref{sa_accuracy_plot} and \ref{sa_time_plot} respectively.

	Evident here is the clear and definite reduction in solution accuracy from 99.6 \ to 72.3 \ as the problem size increases. Also, the execution time quickly grows to be very lengthy at around $ 500 \unit{s} $ with 11 debris (154 binary variables). These plots highlight the problems with using this method to solve large problems, for example with $ N_t $ in the order of 100 debris objects (10,300 binary variables), being that very long execution times and high probability of obtaining an invalid solution should be expected.

	\section{Quantum Annealing}

\begin{figure}[H]
		\centering
		\includegraphics[width=0.55\textwidth]{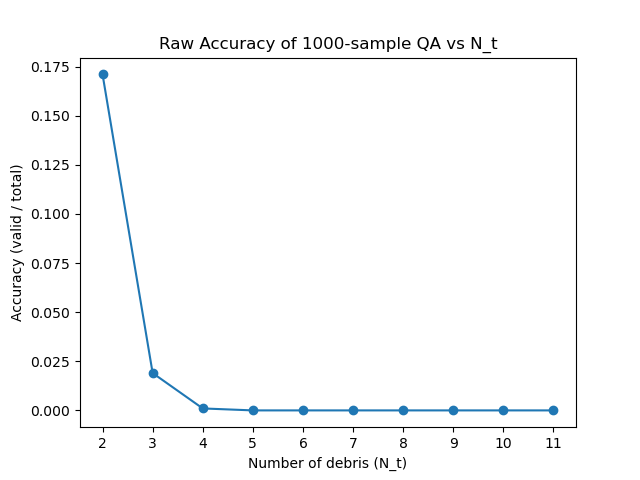}
		\caption{Quantum annealing raw accuracy with varying artificial problem instance size}
		\label{qa_raw_accuracy_plot}
	\end{figure}

	Solution of problems directly using the D-Wave QPU proved to be difficult and was only effective when solving the smallest problems. Figure \ref{qa_raw_accuracy_plot} shows the solution accuracy with anneal time set to $ 50\unit{\mu s} $ as the size of the problem is increased. As can be seen, the accuracy is fairly small at around 17\ with 2 debris (10 binary variables), and sharply drops to almost or exactly 0\ when the number of debris is increased to 4 (28 binary variables) and greater.

	Some investigation in search of the reason for the low solution accuracy reveals that the chaining of qubits to represent variables is relevant. As explained in section \ref{imp_quantum_annealing}, larger problems require an increasing number of physical qubits but the relationship is exponential due to each qubit requiring a number of connected neighbours proportional to the problem size.

\begin{figure}[H]
		\centering
		\includegraphics[width=0.55\textwidth]{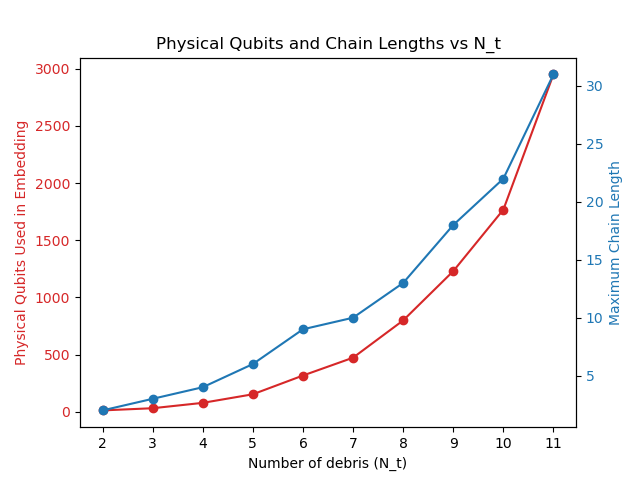}
		\caption{Exponentially increasing physical qubits and chain lengths with growing problem size}
		\label{qa_physical_qubits_chain_lengths}
	\end{figure}

	The length of the longest chain in the embedding of variables as qubits directly follows the number of physical qubits required, as can be seen in Figure \ref{qa_physical_qubits_chain_lengths}. Using simple graph fitting techniques, the physical qubits as a function of $ N_t $ shown on this plot is found to approximately match the exponential expression $ 25e^{0.43{N_t}}-50 $ and the result is that for the 11 debris (154 binary variable) problem, close to 3,000 physical qubits and chains of length greater than 30 are required.

\begin{figure}[H]
		\centering
		\includegraphics[width=0.55\textwidth]{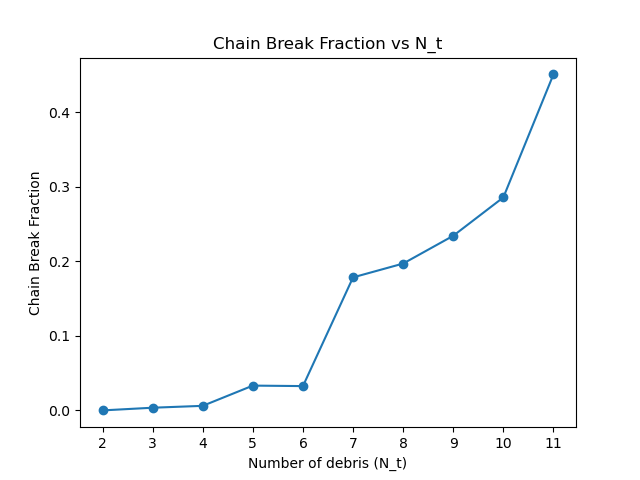}
		\caption{Increasing quantum annealing chain break fraction with growing problem size}
		\label{qa_chain_breaks}
	\end{figure}

	With chains this long the probability of chain breaks (where qubits in a chain provide different measurement values to each other) given any other configuration is very high as can be seen in Figure \ref{qa_chain_breaks}. The chain strength here is automatically calculated by the D-Wave backend appropriately, based on the size of the QUBO coefficients. The chain break fraction becomes higher than 0.4 with 11 debris objects (154 binary variables), which is a significant problem given a broken chain can result in severe degradation of the solutions obtained from the QPU as the logic within the formulation upheld by the qubits is violated.

	Using guidelines from the D-Wave documentation\cite{dwave_chains} on how to manually set chain strength higher to avoid chain breaks, the problems were solved with the chain strength increased to $ 74,200 $ (7 times greater than the largest QUBO coefficent absolute value of $ 10,600 $), but despite the chain break fraction being successfully reduced, the results did not improve and were in fact worse.

	Furthermore, each coupling between a given pair of qubits adds an unwanted bias to each due to integrated control errors (ICE)\cite{dwave_ice} in the QPU, which can be significant with chained qubits\cite{dwave_spin_reversal}. To reduce this error, spin-reversal transforms can be applied to the qubits prior to interpreting the measurement of each qubit. Increasing the number of spin reversal transforms from 1 to 100 increased the QPU access time from $ 16 \unit{ms} $ to $ 424 \unit{ms} $, with the result unreliable in that it could even decrease the solution accuracy severely depending on other parameters.

	Further technical methods were employed in an attempt to increase the solution accuracy as per the advice given in the D-Wave documentation\cite{dwave_config}. This presented further difficulty, mainly because most attempts to increase the accuracy made the execution significantly more expensive in terms of the free time available to use \textit{D-Wave Leap} resources.

\begin{figure}[H]
		\centering
		\includegraphics[width=0.55\textwidth]{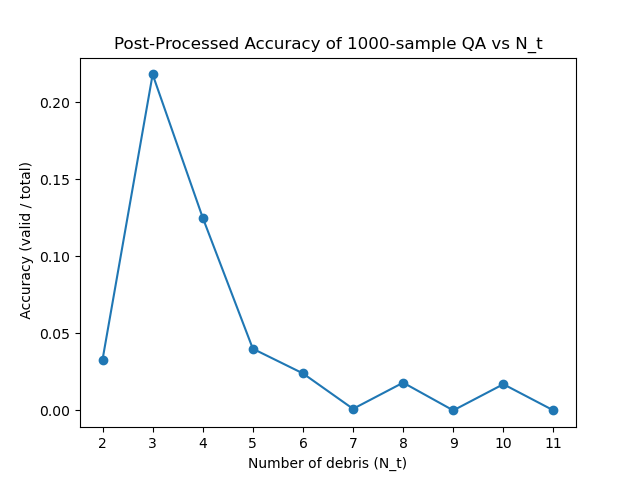}
		\caption{Quantum annealing post-processed accuracy with varying artificial problem instance size}
		\label{qa_pp_accuracy_plot}
	\end{figure}

	However, one technique that clearly helped to obtain a valid answer was post-processing of the samples obtained from the QPU. Figure \ref{qa_pp_accuracy_plot} shows the solution accuracy obtained by supplying the steepest descent algorithm with these samples showing improvements particularly for larger problems. For example, with the number of debris as 8 and 10 (88 and 130 binary variables), this method took a result of 0 valid samples to a small number of valid samples, at least finding a solution to the problem.

	Given the limited time available to access the D-Wave quantum computing resources, the investigation into poor solution accuracy using the QPU directly was stopped here because the clear improvement from introducing a classical element into the processing of results strongly suggests that the use of resources is more well spent by investigating the D-Wave hybrid solvers, as outlined in the next section.

\begin{figure}[H]
		\centering
		\includegraphics[width=0.55\textwidth]{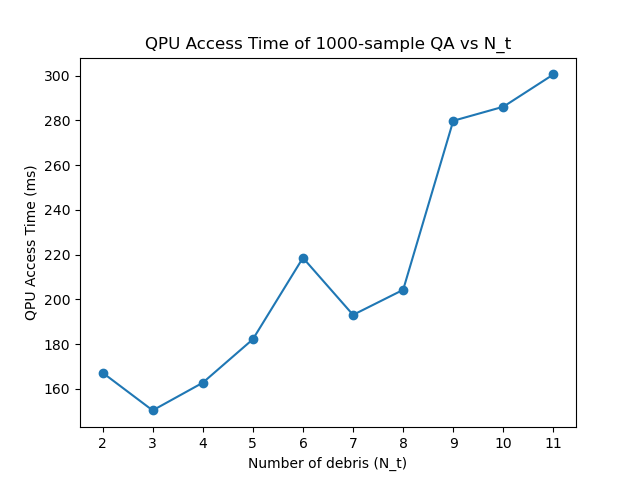}
		\caption{Quantum annealing execution time with varying artificial problem instance size}
		\label{qa_time_plot}
	\end{figure}

	Figure \ref{qa_time_plot} shows the QPU access time with varying number of debris objects. As can be seen, the total access time clearly increases as the number of debris and therefore variables increases. The main contributor to the access time is the sampling time, and since each binary variable representing the problem is implemented using one or more qubits and the final solution is obtained from measuring these qubits, it is no surprise that larger problems take proportionally much longer, simply because more qubits must be measured to obtain each sample.

	It should also be noted that the time taken to prepare the problem definition in code and the data transfer to and from the D-Wave backend became significant as the problem size increases. It is suspected that the data structures used to represent the problem in software could be optimised, as could the hardware used to perform these steps, in order to improve the execution of the algorithm in this regard.

	\section{Quantum-classical Hybrid Algorithm}\label{res_hybrid}

	Impressively, all artificially constructed problems of size 2 through to 11 debris were solved successfully by the D-Wave hybrid solver, taking around $ 3 \unit{s} $ QPU access time to find the optimum solution in every case.

    The format of the answer returned was simply the known minimum energy solution, with no statistical analysis of the data returned required. The total time taken to send the problem data, execute the algorithm and retrieve the results was around $ 6 \unit{s} $, again suspected to be due to extra time required to construct the problem definition locally and transmit data over the internet.

	This result is very promising and is put into practice with a dataset representing a real debris cluster in the next chapter.

	\chapter{Practical Application on Kosmos-1408 Debris Dataset}\label{practical_application}

	Kosmos-1408 was an electronic signals intelligence (ELINT) satellite operated by the Soviet Union, launched into low Earth orbit in 1982. It operated for around two years before becoming inactive and left in orbit, and was destroyed in a Russian anti-satellite weapon test in 2021, resulting in space debris in orbits between 300 and 1,100 km.

	Using \textit{pykep} 2.6.2 developed by the European Space Agency (ESA)\cite{pykep}, a TLE dataset containing orbital information about 79 debris objects from this satellite obtained on 24 August 2023 from CelesTrak was used to create the QUBO model. The quantum-classical hybrid method was then used to select 5 debris objects from this large group for removal in an ADR mission within 365 days, with 20 days service time required by the spacecraft to dispose of each debris object. The reference date for calculating the osculating elements of section \ref{orbital_mechanics} is 30 September 2023, which represents the planned time for transfer to the first debris object. This problem is extremely large, requiring 6,478 binary variables to be represented using the mathematical formulation of this work.

	A valid solution (breaking none of the constraints) was found that uses $ 870 \unit{m} / \unit{s} = 0.87 \unit{km} / \unit{s} $ propellant to perform the inter-debris transfers and disposals using the methods described in section \ref{orbital_mechanics}, in a mission lasting 241 days.

	The ordered list of named debris objects representing this solution is shown below:

	\begin{enumerate}
		\setlength{\itemindent}{4em}
		\item 1982-092RP
		\item 1982-092FT
		\item 1982-092RG
		\item 1982-092BYD
		\item 1982-092JA
	\end{enumerate}

	The breakdown of costs in terms of propellant is as follows:

	\begin{itemize}
		\setlength{\itemindent}{4em}
		\item[--] Transfer cost: 200 m/s
		\item[--] Disposal cost: 670 m/s
		\item[--] Total cost:    870 m/s
	\end{itemize}

	The debris orbital transfer times in days from the reference date are as follows:

	\begin{itemize}
		\setlength{\itemindent}{4em}
		\item[] 1-2: 92
		\item[] 2-3: 126
		\item[] 3-4: 198
		\item[] 4-5: 241
	\end{itemize}

	Solution of this problem required around 25 s QPU access time. However, it took around $ 100 \unit{s} $ to create the \textit{Python} data structures representing the QUBO model and around $ 300 \unit{s} $ to send the problem data, perform the execution and retrieve the results. Optimising the representation of the QUBO model sent to D-Wave Leap so as to minimise this extra time taken would be a worthwhile next step to improve on the work done here.

	This solution is easily verified as valid, in that none of the constraints of the formulation are violated. However, it is not known if it is the optimal solution. To explicitly check every possible solution would require enumeration of $ \frac{N_t!}{(N_t - N_s)!} $ candidates which, with $ N_t = 79 $ and $ N_s = 5 $, is a huge number and a completely unreasonable task even with the powerful classical computers available today. As an illustration, attempting to calculate this value using a common scientific calculator gives ``Math Error".

	The fact that optimality of the solution presented in this chapter cannot be easily verified should not be concerning. On the contrary, the inability to verify the solution should highlight how remarkable it is that a valid solution was found. This demonstrates the incredible strength of quantum computing hardware combined with intuitive classical computational methods in solving this complicated real-world problem.

	\chapter{Conclusion}\label{conclusion}

	A strategy for the analysis of active debris removal missions targeting multiple objects from a set of objects in near-circular orbit with similar inclination has been presented. Algebraic techniques successfully reduce the orbital mechanics regarding specific inter-debris transfer and disposal methods to simple computations, which can then be used as the coefficients of a QUBO problem formulation. This formulation minimises the total propellant used in the mission, constrains the order of debris selection to conform with the method of inter-debris transfer chosen in the strategy with allowances for servicing time and constrains the total mission time to be within a set deadline.

	The modelling of this problem was tested and validated by artificially constructing small problems of varying size (from 2 to 11 debris) and solving these problems using classical computational methods; steepest descent, tabu search and simulated annealing. The main measures of algorithm success considered were the ability to find a valid solution and the execution time. The clear weaknesses in using these methods in isolation were documented and explained, though the application of these algorithms successfully validated the efficacy of the QUBO model constructed to implement the debris selection strategy.

	Having confirmed the reliability of the QUBO model using classical methods, quantum annealing hardware provided by D-Wave was used to solve the artificial problems. Direct execution using a QPU proved difficult and no conclusion was reached regarding the optimality of configuration parameters, but the use of solutions obtained from the QPU as the initial states in the steepest descent algorithm proved to be effective for the larger problems of this artificial set. Use of quantum-classical hybrid solvers was very effective and provided the optimal solution to each problem using around $ 3 \unit{s} $ of QPU access time.

	Finally, the classical-quantum hybrid method was successfully used to solve a large problem using a real dataset. From a set of 79 debris objects resulting from the destruction of the Kosmos-1408 satellite, an active debris removal mission starting on 30 September 2023 targeting 5 debris objects for disposal within a year with 20 days servicing time per object was successfully planned. This plan calculated the total propellant cost of transfer and disposal to be $ 0.87 \unit{km} / \unit{s} $ and would be complete well within the deadline at 241 days from the start date. This problem used 6,478 binary variables in total and was solved using around $ 25 \unit{s} $ of QPU access time.

	Generally, with increasingly large problems the time taken to prepare the data structures in code, send data to and retrieve data from the D-Wave backend leading to lengthy total execution times are the main weakness seen in the results of this work. This could be investigated further and improved in future work following on from this. The focus of this work has been on creating an effective formulation that can be used to solve large real-world active debris removal planning problems using quantum annealing resources in reasonable time, which has been done successfully.

	\bibliographystyle{unsrtnat}
	\bibliography{biblio}

	\appendix
	\chapter{Artificial Problem Construction}\label{appendix_artificial_problems}

	The smallest instance is set up as follows:

	\begin{equation} \label{2_instance_n_t}
	N_t = 2
	\end{equation}

	\begin{equation} \label{2_instance_n_s}
	N_s = 2
	\end{equation}

	\begin{equation} \label{2_instance_t_max}
	T_{max} = 7
	\end{equation}

	\begin{equation} \label{2_instance_t_s}
	T_s = 1
	\end{equation}

	\begin{equation} \label{2_instance_t_ij}
	\textbf{T} = \begin{pmatrix}
					 0 & 2\\
					 2 & 0\\
	\end{pmatrix}
	\end{equation}

	\begin{equation} \label{2_instance_c_ij}
	\textbf{C} = \begin{pmatrix}
					 0 & 1\\
					 1 & 0\\
	\end{pmatrix}
	\end{equation}

	\begin{equation} \label{2_instance_c_i}
	\textbf{c} = \begin{pmatrix}
					 1\\
					 6\\
	\end{pmatrix}
	\end{equation}

	When the total number of debris $ N_t $ is increased to 3, so is $ N_s $:

	\begin{equation} \label{3_instance_n_t}
	N_t = 3
	\end{equation}

	\begin{equation} \label{3_instance_n_s}
	N_s = 3
	\end{equation}

	\begin{equation} \label{3_instance_t_max}
	T_{max} = 7
	\end{equation}

	\begin{equation} \label{3_instance_t_s}
	T_s = 1
	\end{equation}

	\begin{equation} \label{3_instance_t_ij}
	\textbf{T} = \begin{pmatrix}
					 0 & 2 & 4\\
					 2 & 0 & 6\\
					 4 & 6 & 0\\
	\end{pmatrix}
	\end{equation}

	\begin{equation} \label{3_instance_c_ij}
	\textbf{C} = \begin{pmatrix}
					 0 & 1 & 3\\
					 1 & 0 & 2\\
					 3 & 2 & 0\\
	\end{pmatrix}
	\end{equation}

	\begin{equation} \label{3_instance_c_i}
	\textbf{c} = \begin{pmatrix}
					 1\\
					 6\\
					 1\\
	\end{pmatrix}
	\end{equation}

	When the total number of debris $ N_t $ is increased to 4 or greater, $ N_s $ remains at 3:

	\begin{equation} \label{4_instance_n_t}
	N_t = 4
	\end{equation}

	\begin{equation} \label{4_instance_n_s}
	N_s = 3
	\end{equation}

	\begin{equation} \label{4_instance_t_max}
	T_{max} = 7
	\end{equation}

	\begin{equation} \label{4_instance_t_s}
	T_s = 1
	\end{equation}

	\begin{equation} \label{4_instance_t_ij}
	\textbf{T} = \begin{pmatrix}
					 0 & 2 & 4 & 7.1\\
					 2 & 0 & 6 & 7.1\\
					 4 & 6 & 0 & 6\\
					 7.1 & 7.1 & 6 & 0\\
	\end{pmatrix}
	\end{equation}

	\begin{equation} \label{4_instance_c_ij}
	\textbf{C} = \begin{pmatrix}
					 0 & 1 & 3 & 0.5\\
					 1 & 0 & 2 & 0.5\\
					 3 & 2 & 0 & 3\\
					 0.5 & 0.5 & 3 & 0\\
	\end{pmatrix}
	\end{equation}

	\begin{equation} \label{4_instance_c_i}
	\textbf{c} = \begin{pmatrix}
					 1\\
					 6\\
					 1\\
					 2\\
	\end{pmatrix}
	\end{equation}

	For problem instances with $ N_t \ge 4 $, the only valid solutions (satisfying all of the 8 constraints) and their corresponding energies are shown in Table \ref{toy_problem_solutions}. For the smallest problem, the optimum solution is trivial as $ N_t = N_s = 2 $ and for problems with $ N_t = 3 $ the solution from the table including node 4 is obviously excluded.

	\begin{table}[H]
		\centering
		\begin{tabular}{||c | c||}
			\hline
			Solution & Energy \\ [0.5ex]
			\hline\hline
			1,3,4 & 10 \\
			\hline
			1,2,3 & 11 \\
			\hline
			2,1,3 & 12 \\
			\hline
			1,3,2 & 13 \\ [1ex]
			\hline
		\end{tabular}
		\caption{Artificial problem instance solutions}
		\label{toy_problem_solutions}
	\end{table}

\begin{figure}[H]
		\centering
		\begin{tabular}{cc}
			\includegraphics[width=0.45\textwidth]{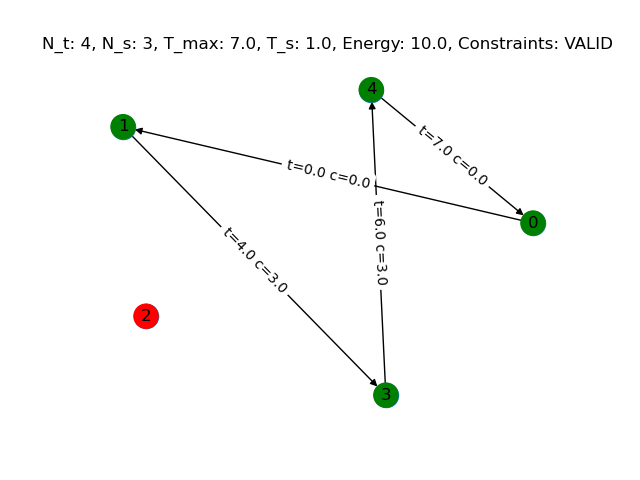} &   \includegraphics[width=0.45\textwidth]{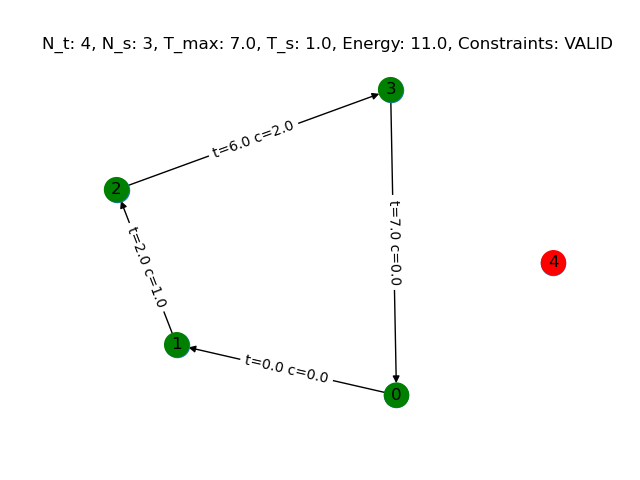} \\
			\includegraphics[width=0.45\textwidth]{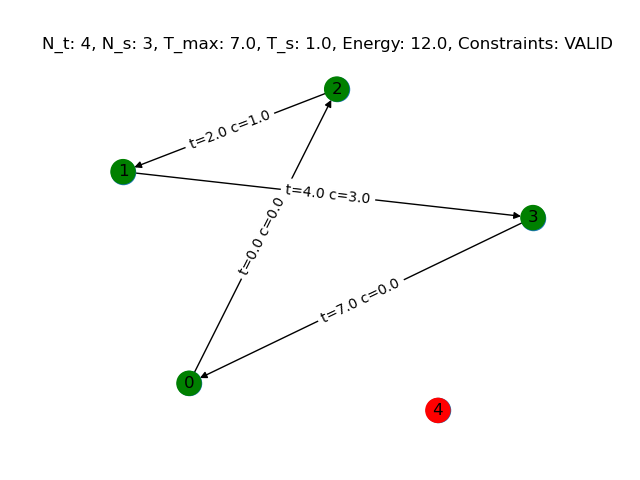} &   \includegraphics[width=0.45\textwidth]{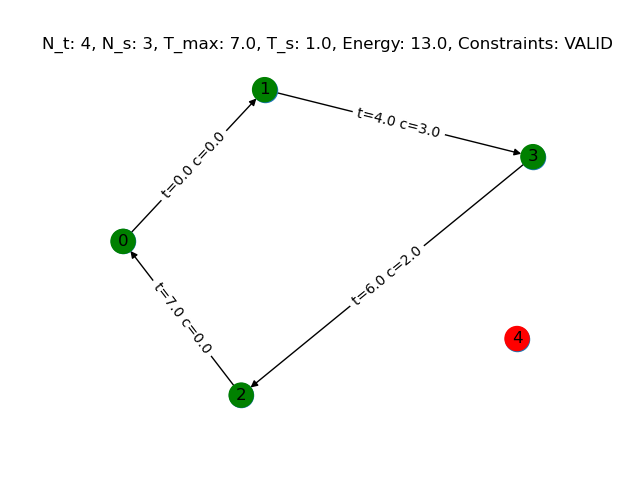} \\
		\end{tabular}
		\caption{Graphs showing solutions to small problem instances}
		\label{toy_solution_graphs}
	\end{figure}

	The solutions for $ N_t = 4 $ are shown in graphical form in Figure \ref{toy_solution_graphs}

	Larger problem instances that have the same solutions are also examined. For example, with $ N_t = 6 $ the set up is as follows:

	\begin{equation} \label{large_instance_n_t}
	N_t = 6
	\end{equation}

	\begin{equation} \label{large_instance_n_s}
	N_s = 3
	\end{equation}

	\begin{equation} \label{large_instance_t_max}
	T_{max} = 7
	\end{equation}

	\begin{equation} \label{large_instance_t_s}
	T_s = 1
	\end{equation}

	\begin{equation} \label{large_instance_t_ij}
	\textbf{T} = \begin{pmatrix}
					 0 & 2 & 4 & 7.1 & 7.1 & 7.1\\
					 2 & 0 & 6 & 7.1 & 7.1 & 7.1\\
					 4 & 6 & 0 & 6 & 7.1 & 7.1\\
					 7.1 & 7.1 & 6 & 0 & 7.1 & 7.1\\
					 7.1 & 7.1 & 7.1 & 7.1 & 0 & 7.1\\
					 7.1 & 7.1 & 7.1 & 7.1 & 7.1 & 0\\
	\end{pmatrix}
	\end{equation}

	\begin{equation} \label{large_instance_c_ij}
	\textbf{C} = \begin{pmatrix}
					 0 & 1 & 3 & 0.5 & 10 & 10\\
					 1 & 0 & 2 & 0.5 & 10 & 10\\
					 3 & 2 & 0 & 3 & 10 & 10\\
					 0.5 & 0.5 & 3 & 0 & 10 & 10\\
					 10 & 10 & 10 & 10 & 0 & 10\\
					 10 & 10 & 10 & 10 & 10 & 0\\
	\end{pmatrix}
	\end{equation}

	\begin{equation} \label{large_instance_c_i}
	\textbf{c} = \begin{pmatrix}
					 1\\
					 6\\
					 1\\
					 2\\
					 10\\
					 10\\
	\end{pmatrix}
	\end{equation}

	It should be noted that the elements of $ \textbf{T} $ added are all equal to $ 7.1 $, which forces all other candidate solutions introduced by the increase in $ N_t $ from $ 4 $ to $ 6 $ to automatically break constraint 8 since the mission deadline $ T_{max} = 7 $. Similarly, elements added to $ \textbf{C} $ and $ \textbf{c} $ are equal to $ 10 $, which means further candidate solutions have a much higher penalty in equation \ref{minimise_cost} since the propellant costs of transfer and disposal are higher. This strategy keeps the valid solution set the same throughout the testing.

\end{document}